\documentclass[a4paper,fleqn,usenatbib]{mnras}
\usepackage{newtxtext,newtxmath}

\usepackage[T1]{fontenc}
\usepackage{ae,aecompl}

\usepackage{graphicx}	% Including figure files
\usepackage{amsmath}	% Advanced maths commands
\usepackage{amssymb}	% Extra maths symbols
\usepackage{mathtools}
\usepackage{hyperref}

\newcommand{\note}[1]{\textcolor{black}{#1}}

\title[Magnetic fields in disc galaxies]{The origin of the structure of large--scale magnetic fields in disc galaxies}

\author[Nixon et al.]{
C. J. Nixon$^{1,2}$\thanks{E-mail: \href{mailto:cjn@leicester.ac.uk}{cjn@leicester.ac.uk}},
T. O. Hands$^3$,
A. R. King$^{1,4,5}$ and
J. E. Pringle$^{1,6}$\\
$^{1}$Theoretical Astrophysics Group, Department of Physics \& Astronomy, University of Leicester, Leicester, LE1 7RH, UK\\
$^{2}$Kavli Institute for Theoretical Physics, University of California, Santa Barbara, CA 93106, USA\\
$^{3}$Institut f\"ur Computergest\"utzte Wissenschaften, Universit\"at Z\"urich, Winterthurerstrasse 190, 8057 Z\"urich, Switzerland\\
$^{4}$Anton Pannekoek Institute, University of Amsterdam, Science Park 904, 1098 XH Amsterdam, Netherlands\\ 
$^{5}$Leiden Observatory, Leiden University, Niels Bohrweg 2, NL-2333 CA Leiden, Netherlands\\
$^{6}$Institute of Astronomy, Madingley Road, Cambridge, CB3 0HA, UK
}
  
\date{Draft version, \today.}
\pubyear{2017}
\pagerange{\pageref{firstpage}--\pageref{lastpage}}
\begin{document}
\label{firstpage}
\maketitle

\begin{abstract}
The large--scale magnetic fields observed in spiral disc galaxies are often thought to result from dynamo action in the disc plane. However, the increasing importance of Faraday depolarization along any line of sight towards the galactic plane suggests that the strongest polarization signal may come from well above ($\sim$0.3--1\,kpc) this plane, from the vicinity of the warm interstellar medium (WIM)/halo interface. We propose \citep[see also][]{Henriksen:2016aa} that the observed spiral fields (polarization patterns) result from the action of vertical shear on an initially poloidal field. We show that this simple model accounts for the main observed properties of large--scale fields. We speculate as to how current models of optical spiral structure may generate the observed arm/interarm spiral polarization patterns.
\end{abstract}

\begin{keywords}
(magnetohydrodynamics) MHD --- dynamo --- galaxies: magnetic fields --- polarization
\end{keywords}

\section{Introduction}
\label{sec:intro}
Large--scale magnetic fields are found in all disc galaxies by observing radio synchrotron emission. Recent reviews of the properties of these fields are given by \cite{Beck:2013aa} and \cite{Beck:2016aa}. Analysis of the polarization properties of the radio emission shows that ordered fields with spiral patterns are found in grand--design, barred and flocculent galaxies. 

Radio polarization maps reveal three characteristic features.\\
(i) Spiral field patterns are seen in almost all galaxies at wavelengths $\lambda \sim 3 - 6$\,cm, where Faraday depolarization by the halo is negligible. They are seen even in galaxies lacking (strong) optical spiral structure -- for example NGC 4736 \citep{Chyzy:2008aa,Beck:2013aa}.\\
(ii) Poloidal ``X-shaped'' fields are seen in edge--on galaxies. An example is NGC 891 \citep{Krause:2009aa,Beck:2013aa}, where they are seen in the halo, above the H$\alpha$ disc (warm interstellar medium, WIM). Similar results have been found by the CHANG-ES survey \citep{Irwin:2012aa,Irwin:2012ab,Irwin:2013aa}. Further \cite{Terral:2017aa} provide evidence that the magnetic field in the Galactic halo has the required field geometry to create an X-shape when viewed edge-on from the outside.\\
(iii) As reported by \cite{Braun:2010aa}, \cite{Beck:2012aa} and others, at $\lambda \approx 20$ cm there is a striking asymmetry of the polarized emission along the major axes of spiral galaxies with inclinations less than $60^\circ$. The emission is always almost completely depolarized around the side of the major axis which is kinematically receding. 

\cite{Braun:2010aa} also note that in strongly inclined galaxies both sides of the major axis become depolarized at 20 cm, where most of the polarized emission from the disc is Faraday--depolarized and the emission from the halo dominates. The asymmetry is still visible at 11 cm, but disappears at smaller wavelengths.
 
The properties of the fields, including their overall structure, are usually taken to indicate field amplification and ordering by a magnetic dynamo mechanism. \cite{Beck:2013aa}, \cite{Beck:2012aa} and \cite{Beck:2016aa} discuss the interpretation of these structures in terms of magnetic dynamo theory. Their general conclusion is that the global field patterns observed currently agree with the predictions of the $\alpha$-$\Omega$ dynamo: large--scale spiral patterns are consistent with a superposition of azimuthal dynamo modes, which are signatures of regular fields generated by mean--field dynamos.\footnote{\cite{Tabatabaei:2016aa} have shown that the average magnetic field strength correlates with galaxy rotation speed, but not with $\Omega$ or $\Omega^\prime$. They conclude therefore that the field cannot be due to an $\alpha$-$\Omega$ dynamo.}

In this paper we question whether dynamo action is necessary to produce the observed large--scale magnetic structures. There is no doubt that the conditions for dynamo action exist close to the disc plane, where feedback from the action of star formation and supernovae is most dominant. In the Milky Way, we observe from close to this plane, and the magnetic field structure appears to be complicated -- for example there is ample evidence for field reversals \citep{Frick:2001aa,Han:2006aa,Van-Eck:2011aa} -- which might be suggestive of dynamo action. Recently \cite{Ordog:2017aa} have explored the magnetic field reversal in the Sagittarius-Carina arm. The usual model for the reversal has the reversal plane (i.e. the putative current sheet) perpendicular to the disk plane. However, here they find that this plane is actually not perpendicular, but is in fact almost aligned (to within 30 degrees) with the disk plane -- See their Figure 2. Thus they find $\partial B_\phi/\partial z \gg \partial B_\phi/\partial R$. In any case, the large--scale fields in other galaxies are observed from outside, and are simpler. So here we ask whether these large--scale fields might have a more straightforward origin than the existence of a galactic dynamo.

In Section~\ref{sec:dynamo}, we discuss the standard disc dynamo. This basic model produces only horizontal magnetic fields ($B_R, B_\phi$) within the warm interstellar medium (WIM) which exists in a layer $|z| \le 500$ pc. Although the growth timescales for the models are locally short, the timescales needed to produce a {\it global} ($\sim 10$ kpc) field structure are comparable with the age of the galaxy. Further, producing global poloidal fields (with $B_z \sim B_R$) needs more assumptions. We propose here instead that the global poloidal field is present {\em ab initio}, as a result of the galaxy formation process. This kind of idea was suggested by \cite{Sofue:1987aa}. Recent cosmological simulations \citep{Pakmor:2017aa} find global halo fields with strengths around $1$-$2\,\mu\,$G.

In Section~\ref{sec:simple}, we further propose \citep[see also][]{Henriksen:2016aa} that starting from an initial large--scale poloidal field threading the halo, WIM and galactic disc, shearing motions naturally generate fields with most of the properties of the observed  large--scale ones. We note that this is a toy model for our expectations of the field structure. Clearly this will not explain everything as the dynamo process can contribute particularly at small $z \lesssim z_{\rm WIM}$, but we argue that these ideas constitute an important component of a complete galaxy magnetic field model. In Section~\ref{sec:polarflux} we estimate the polarized flux from this simple model of the large-scale galaxy magnetic field. In Section~\ref{sec:spiral} we speculate how these simple ideas, which generate axisymmetric field structures,  might also explain the observed non--axisymmetric polarization structures in spiral galaxies, where the polarization degree correlates with the structure of the optical spiral arm pattern. This is often interpreted as a result of azimuthally dependent dynamo activity, but we note that the dynamo growth timescales (of order a galactic rotation period) are marginal for this interpretation. We discuss our current understanding of spiral arm formation, and suggest instead that the observed correlation between the magnetic and optical structures results mainly because arm--generated turbulence (feedback from star formation/supernovae) produces a less ordered local field, and so less polarization.  In Section~\ref{sec:discussion} we present a discussion and our conclusions.

\section{The Standard Galactic Dynamo Model}
\label{sec:dynamo}
In its simplest form, the basic assumption of galactic dynamo theory (see, for example, \citealt{Ruzmaikin:1988aa}) is that the magnetic fields are predominantly generated and contained within the warm interstellar medium (the WIM). Typical properties of the WIM gas are a mean temperature $T \sim 10^4$ K, mean electron density $n_e \sim 0.3$ cm$^{-3}$ and to satisfy hydrostatic equilibrium, a vertical scaleheight $H \approx 500$ pc. The medium is assumed to be turbulent because of the action of star formation and/or supernovae. 

The dynamo model uses the thin disc approximation $H \ll R$, assuming that radial and vertical derivatives obey $\partial / \partial R \ll \partial/ \partial z$ \citep{Chamandy:2013aa}. Since the mean field satisfies div ${\bf B} = 0$, this implies that $B_z \ll B_R, B_\phi$. 

\subsection{Axisymmetric models}
In the simple axisymmetric form, \cite[Chapter VII]{Ruzmaikin:1988aa} the field is written in terms of two functions $A$ and $B$, where  $B = B_\phi$, the azimuthal component of the magnetic field, and $A = A_\phi$ which is the $\phi$-component of the vector potential. Thus
\begin{equation}
B_R = - \frac{\partial A}{\partial z},
\end{equation}
and
\begin{equation}
B_z = \frac{1}{R} \frac{\partial}{\partial R} (R A),
\end{equation}
so that $B_z \sim (H/R) B_R$.

Then the equations governing the evolution of the dynamo are
\begin{equation}
\label{Adynamo}
\frac{\partial A}{\partial t} = \alpha_D B + \eta_T \frac{\partial^2 A}{\partial z^2},
\end{equation}
and
\begin{equation}
\label{Bdynamo}
\frac{\partial B}{\partial t} = - R \Omega^\prime \frac{\partial A}{\partial z} + \eta_T \frac{\partial^2 B}{\partial z^2}.
\end{equation}

The first equation generates poloidal field (here $B_R$) using the standard dynamo $\alpha$--effect assumed to be caused by cyclonic turbulence (\citealt{Steenbeck:1966ab}; \citealt{Steenbeck:1966aa}; \citealt{Parker:1979aa}, Chapter 18). The second generates toroidal field from radial field using shear. In both equations field is dissipated by (turbulent) magnetic diffusivity on the time-scale $\sim H^2/\eta_T$. Note that these are {\bf local} equations, in that there is no radial dependence. Thus we only need boundary conditions at $z = \pm H$. The usual assumption for the boundary conditions is that 
\begin{equation}
B_R(\pm H) = B_\phi(\pm H) = 0.
\end{equation}

When solved, the solutions give a value of $B_R/B_\phi$ which is axisymmetric, and hence we get predictions \note{for the field structure that consist of axisymmetric spirals}, with pitch angles essentially depending on the balance between  $\alpha_D$ and the rate of shear, provided that the dissipation caused by $\eta_T$ is not too great. 

\subsection{Timescales}
\label{sec:timescales}
The growth time for the dynamo is given by the geometric mean of the timescales for dynamo action (due to $\alpha_D$) in Equation~\ref{Adynamo} and galactic shear ($R \Omega^\prime$) in Equation~\ref{Bdynamo}. The quantity $\alpha_D$ is not easy to compute, but if dynamo action is operative, then since in the observed galaxies the magnetic pitch angles are of order unity (thus to a first approximation $B_R \sim B_\phi$) it is necessary, and therefore assumed, that these timescales are approximately the same. Thus we estimate the growth rate for the dynamo to be of order the galactic rate of shear. We conclude that, if $\alpha_D$ has the required properties, then the equations can give an exponentially growing field configuration, with the growth timescale typically equal to the galactic rate of shear (or, approximately, the galactic rotation rate). Since a typical galaxy can have rotated around 30 -- 50 times in its lifetime (for example the orbital period of the Sun through the Galaxy is $\sim 200$ Myr and the Galaxy's age is around 10 Gyr), this gives ample time to magnify quite small initial seed fields. 

The dissipative terms depend on the effective magnetic diffusivity $\eta_T$. Thus the decay timescales for such fields can be estimated from the turbulent galactic magnetic diffusivity, generally taken to be
\begin{equation}
\eta_T \sim \frac{1}{3} ul,
\end{equation}
where $u \approx 10\, {\rm km\,s^{-1}}$ is a typical turbulent velocity and $l \approx 50$ pc a typical turbulent lengthscale \citep{Ruzmaikin:1988aa,Shukurov:2004aa}. The decay timescale for magnetic fields within a galactic disc is then
\begin{equation}
\tau_D \sim \frac{H^2}{\eta_T},
\end{equation}
where $H \sim 500$\,pc is the disk scaleheight for the WIM. At around a galactic radius $R = 10$\,kpc this gives $\tau_D \approx 0.5 - 0.7 $ Gyr. This is somewhat longer than the expected dynamo growth timescale of around 0.2 Gyr. It is also about $1/20$ of a galactic lifetime of $\sim 10$\,Gyr \citep{Ruzmaikin:1988aa,Shukurov:2004aa}. These authors suggest that such a short field decay timescale implies that dynamo action is needed to generate the observed magnetic field structures.

However, this growth timescale is not the timescale for the generation of a {\em large--scale} galactic field (for example, \citealt{Arshakian:2009aa}). The dynamo equations~\ref{Adynamo} and~\ref{Bdynamo} have no radial dependence. So the field generated by this mechanism is local to a single disc annulus and generates a field on a scale--length  $\sim H$. Generating a field with a lengthscale of a galactic radius $R$ requires radial communication of some kind. In the standard dynamo equations \citep[Chapter VII]{Ruzmaikin:1988aa} this occurs on a radial magnetic diffusion timescale of order $\tau_R \sim R^2/\eta_T$ which \cite[see below]{Ruzmaikin:1988aa} estimate as 500 Gyr, much longer than the age of the galaxy.  However, \citet[][\note{see also \citealt{Rodrigues:2015aa}}]{Arshakian:2009aa} note that in fact the radial propagation may occur through a combination of radial diffusion (due to $\eta_T$) and local growth (on a timescale $\gamma \sim \Omega$). This generates a propagating magnetic front \citep{Moss:1998aa} which moves at a speed $V \approx 2 \sqrt{\gamma \eta_T}$. The front propagation timescale is $\tau_F \sim R/V \sim \sqrt{\tau_R / \Omega} \note{~\sim 5 - 10}$ \note{Gyr, which is comparable to the age of the galaxy \citep[cf.][]{Rodrigues:2015aa}. Thus the standard galactic dynamo model is capable of generating a large-scale ($\sim R$) magnetic field within the galactic disc, but takes approximately the lifetime of the galaxy to do so.}

\subsection{Vertical Magnetic Field}
The basic dynamo model described above does not permit the generation of large-scale vertical fields $B_z$ threading the disc -- indeed it assumes that such fields do not exist. The fields generated are primarily parallel to the galactic plane and contained within the turbulent WIM disc ($-500$\,pc $\le z \le 500$\,pc). Thus these models need further adjustment and assumptions in order to explain the large-scale poloidal X-shaped fields seen in many edge-on spiral galaxies, and other indications that such large-scale vertical fields exist (see Section 2.4).

\citet[see also \citealt{Henriksen:2017aa}]{Brandenburg:1992aa} postulate the existence of an additional dynamo in the halo, with comparable dynamo parameters to that in the disc, and with a strong magnetic diffusivity $\eta_h = (1/3) u_h l_h$ caused by turbulence in the halo medium with canonical eddy size $l_h \approx 500$ pc and eddy velocity $u_h \approx 100$ km/s. Thus these eddies have a turnover timescale of $\tau_h \sim l_h/u_h \sim 5$ Myr. The timescale for diffusion of flux through the assumed halo size of $R = 15$ kpc is then $\approx R^2/\eta_h \sim 1.5$ Gyr; this is presumably approximately equal to the growth timescale of the halo dynamo. No explanation is given for what powers this turbulence. Since the turbulence is trans-sonic for a standard halo temperature of $\approx 10^6$ K (see Section 2.4), the decay timescale of the turbulence is of order the eddy turnover time of $\approx$ 5 Myr, which is much less than the dynamical timescale for the halo as a whole.

In addition to the postulated halo dynamo, a number of authors \citep{Brandenburg:1993aa,Moss:2008aa,Moss:2010aa,Chamandy:2015aa} include the hypothesis of a galactic wind driven from the top of the WIM\footnote{\note{The idea that the WIM loses mass vertically was introduced by \cite{Shukurov:2006aa} in order to remove magnetic helicity. The aim of this assumption was to quench dynamo action within the WIM and thus determine limiting values of $B_R$ and $B_{\phi}$ within the WIM. They did not consider the generation of a global $B_z$ field, encompassing the halo. They postulate a mass {\em loss} rate of $\dot{M} = 1.5 M_\odot$/yr by assuming that it is the same as the mass {\em circulation} rate within a ``galactic fountain'' flow described, for example, by \cite{Shapiro:1976aa}. This assumed net mass {\em loss} is also in contrast to the usual assumption of the need for a net mass {\em gain} by the galactic disk of a few $M_\odot$/yr to satisfy the constraints of the chemical evolution of disk galaxies \citep[see, for example,][]{Lacey:1985aa}}}. The wind is designed to advect large-scale field; \cite{Brandenburg:1993aa} suggest that the field is generated in the disc and then distorted by the wind in the halo. It is suggested that the wind is driven by supernovae and star formation in the galactic disc. Such winds are indeed observed in strongly star forming galaxies \citep[see the review by][]{Heckman:2017aa}, and are driven from the disc with typical outflow velocities of $500-1$,$000$\,km/s, comparable to the escape velocity from the dark matter halo. However, this contrasts with the wind properties required by the dynamo models which have velocities at the disc in the range $1-3$\,km/s, only reaching larger velocities of $50-200$\,km/s at the outer radius of the halo at $15-20$\,kpc. No explanation is given of how such a wind is generated, or accelerated, especially given that the acceleration process must occur throughout the halo, and that the wind does not reach the galactic escape velocity.\footnote{\cite{Moss:2010aa} are aware of the discrepancy. They note that a typical galactic wind does indeed have a velocity of some hundreds of km/s, and suggest that the slow outflow postulated to be relevant to the dynamo models must ``correspond to the velocity of a component that comprises a tiny fraction of the galactic medium''.  What this component might be, or why it fails to interact dynamically or magnetically with the rest of the halo medium, is not made clear.}

We argue here that there may be no need to invoke disc dynamo action to explain global field structures, and that these may instead arise simply because disc galaxies have large--scale poloidal fields from early times. So we suggest here that a global {\it poloidal} field threads both the WIM and the halo. \citet[Chapter V]{Ruzmaikin:1988aa} dismiss this possibility and argue that as the protogalactic gas collapses to form a contracting disc, any such field would form a magnetic pinch in the disc plane which would reconnect at the Alfv\'en speed. They conclude that there should be no such global field threading the disc.  In contrast, \cite{Sofue:1987aa} argue that a global poloidal field could be primordial because it  could only diffuse out of the galactic plane on a timescale of order $\sim R^2/\eta_T \sim 200-350$\,Gyr. They suggest that such a global field, sheared by the rotation in the disc plane, could also produce a spiral magnetic pattern. 

In fact neither of these estimates of poloidal field decay is correct. As shown by \cite{van-Ballegooijen:1989aa} (see also \citealt{Lubow:1994aa}) the correct estimate for the decay of such a poloidal field threading a disk is
\begin{equation}
\tau_D \sim \frac{RH}{\eta_T},
\end{equation}
which gives a decay timescale of order 10\,Gyr, comparable to the age of a galaxy. This implies that as \cite{Sofue:1987aa} suggest, galaxies may well be threaded by global poloidal magnetic fields which they acquired at the time of their formation.

\subsection{Faraday depolarization}
\label{FD}
The presence of a large--scale poloidal field means that we must take account of Faraday rotation when interpreting polarization patterns in terms of large--scale magnetic structure. There are two main problems here. First, we can only see polarization where there is both a large--scale organized field {\em and also} enough relativistic electrons (cosmic rays) to generate the observed synchrotron emission. Second, we deduce the field structure from the polarization directions \note{(combined with Faraday rotation measures)} of the synchrotron emission, typically at $\lambda = 3 - 20$\,cm, but in a magnetic medium -- here, one with  a large--scale poloidal field --  the polarization direction is affected by Faraday rotation. 

To get some idea of the numbers involved, we suppose that the WIM, with properties mentioned above, is sandwiched between hotter and less dense gas which we will call the ``halo''. Following \cite{Ferriere:2001aa}, we take the halo to have temperature $T \sim 10^6$\,K and electron density $n_e \sim 0.003$\,cm$^{-3}$. Then hydrostatic equilibrium gives the halo scaleheight as $H \sim 5$\,kpc $\sim R$. Note that we have chosen the thermal pressure in the WIM and the halo to be the same, at approximately $n_e T \sim 3000$ K/cm${^3}$. For a field in equipartition with the thermal pressure we have
\begin{equation}
  \label{equiB}
B \approx 4 \left( \frac{nT}{3000} \right)^{1/2} \mu{\rm G}.
\end{equation}

The angle through which the apparent polarization of a radio wave is rotated is
\begin{equation}
\Phi = {\rm FD} \lambda^2,
\end{equation}
where $\lambda$ is the wavelength in metres and FD is the Faraday polarization parameter. 
FD is defined as
\begin{equation}
  \label{fara}
{\rm FD} = 0.8 \int n_e \, B_\parallel \, dl,
\end{equation}
where the integral is along the line of sight and the units imply that $n_e$ is in cm$^{-3}$, $B_{\parallel}$ is the component of magnetic field (in $\mu$G)  in this direction, and $dl$ is in pc (see, for example, \citealt{Beck:2016aa}).

For each wavelength, there is a critical value FD$_{\rm cr}$ of the Faraday polarization parameter which would give $\Phi \approx 1$. So FD$_{\rm cr}$ specifies the distance into the medium to which we can observe polarized flux. We note that (using FD$_{\rm cr} \lambda^2 = 1$ where $\lambda$ is in metres) for $\lambda = 6$ cm, FD$_{\rm cr}$ $\approx$ 280, and for $\lambda = 20$ cm, FD$_{\rm cr}$ $\approx 25$. For the estimated parameters (given above) of the WIM disc, and of the halo, we use FD $\approx n_e B H$ to find FD(Halo) $ \approx 48$, and FD(WIM) $\approx 480$. Thus at 20 cm we would expect to see polarization roughly to the bottom of the halo, whereas at 6 cm we can see into, but not necessarily through, the WIM disk.

Observing at two wavelengths gives a measure of the effect of Faraday rotation. The maximum values of FD measured at 20 cm are typically around $\pm$ 20 (\citealt{Mao:2015aa}, M51; \citealt{Beck:2015aa}, IC 342; \citealt{Beck:2007aa}, NGC 6946), and at 6 cm the maximum measured values are typically around $\pm$ 100 (\citealt{Fletcher:2011aa}, M51; \citealt{Beck:2007aa}, NGC 6946). Thus the maximum measured values typically lie close to the critical values. This suggests that the full FD through the galaxy is larger than these values, and that we can only ever ``see'' the field structure through polarization down to a depth corresponding to FD $\approx$ FD$_{cr}$.

Indeed, as noted by \citet[IC243]{Beck:2015aa}: ``The disks of IC 342 and many other galaxies studied so far are not transparent to polarized decimeter radio waves; they are {\em Faraday thick}.'' In their study of M51, \cite{Fletcher:2011aa} find that the patterns of polarization angle at $\lambda =  3$, 6 cm and $\lambda = 18$, 20 cm are very different. They explain this by invoking two distinct Faraday rotating layers at heights 1 -- 2 kpc above the plane. They do argue, however, that the disc is Faraday thin at 6 cm. In addition, \cite{Mao:2015aa} require a coherent vertical magnetic field in M51's halo as well as the plane-parallel component, and estimate the WIM scaleheight in M51 to be at least 1.2 kpc. To interpret the observed polarization structures in terms of magnetic fields in the plane of the disc, and so as evidence for magnetic dynamo action, it is essential that the galactic plane should be Faraday {\it thin} at the relevant wavelength (typically around 6 cm). As we have seen, this assumption is marginal. 

\section{A Simple model for large--scale axisymmetric spiral fields}
\label{sec:simple}
With these ideas in mind, we now investigate a simple physical model, close to that proposed by \cite{Sofue:1987aa}. We note that these ideas have also been investigated more recently by \cite{Henriksen:2016aa}.  Our aim is to see if the simple assumption of a galaxy initially threaded by a large--scale poloidal field can give a straightforward explanation of the three basic observational properties of large scale fields mentioned in the Introduction, namely:

\begin{enumerate}
\item almost all disk galaxies, including those with little or no evidence of optical spiral arms, have spiral magnetic field patterns,

\item edge--on disc galaxies have large--scale X-shaped poloidal magnetic field structures,

and 

\item the polarized 20\,cm emission is always almost completely depolarized around that side of the major axis which is kinematically receding.
\end{enumerate}

\subsection{Initial conditions}
\label{ICs}
We take as our starting point a simple model of gas in a galaxy. Given that we may only be able to see the magnetic field structure down to a height $\sim 0.3 - 1$\,kpc above the disc plane, the only gas components relevant for the evolution of global magnetic fields are the WIM and the halo (or corona). 

We take the background galaxy potential to be fixed and we ignore cold neutral or molecular gas, which comprises the star--forming layer close to the plane ($|z| \lesssim 100$\, pc). We consider WIM ($10^4$ K) and halo ($10^6$ K) gas in pressure equilibrium with the properties given above \citep{Ferriere:2001aa}. We take the scaleheight of the WIM to be 500\,pc and the scaleheight of the halo to be 5\,kpc.

We note that because of the difference in scaleheights, there is also a difference in radial pressure gradients, which in turn implies that there must be a vertical shear ($d \Omega/dz$), so that the two components rotate at different angular velocities. The difference in angular velocities is $\sim \Delta \Omega/\Omega \sim (H/R)^2 \sim 0.1 - 0.2$, corresponding to a velocity difference $\Delta V \sim 20 - 40\, {\rm km\, s^{-1}}$. There is ample observational evidence for vertical shear. \cite{Swaters:1997aa} in NGC 891 find $\Delta V \sim 25\, {\rm km\, s^{-1}}$ over a height of $\sim 2$\,kpc. In NGC 2403 \cite{Schaap:2000aa} find a thin disk, and a thicker $1 - 3$\,kpc HI layer which rotates more slowly with $\Delta V \sim 25\,{\rm km\, s^{-1}}$. From optical emission lines (H $\alpha$, [NII], [SII]) \cite{Fraternali:2001aa} find that the ionized gas rotates more slowly than the disc, with $\Delta V \sim 20 - 50\, {\rm km\, s^{-1}}$, and also a suggestion of a net radial inward flow with $V_R \sim 10 - 20\, {\rm km\, s^{-1}}$. More recently \cite{Zschaechner:2015aa,Zschaechner:2015ab} have detected similar vertical shear in several galaxies.

Given this, we assume an initially vertical magnetic field threads the protogalactic disc. We expect that the gas motions will shear this initial field to give a spiral pattern as seen from above, and also the asymmetry along the major axis. To give ``X-shaped'' poloidal fields we need some radial flow. We expect this be generated by the vertical shear, since with a poloidal field this causes angular momentum transport, and hence radial flow. We note that between the WIM and the halo we have a difference of a factor $\sim 100$ in number density $n$, so pressure balance requires that the  temperature $T \propto n^{-1}$. The scaleheights vary as $H \propto T^{1/2}$, implying surface densities $\Sigma \propto T^{-1/2}$, and so the halo mass is $\sim 1/10$ of the WIM mass. This implies that the halo is spun outwards, suggesting an origin for the X--shaped poloidal field. We explore these ideas in a numerical model in the next Section, and also note that similar results have been obtained by \cite{Henriksen:2016aa}.

We also need a source of synchrotron (cosmic ray) electrons to explain the visibility of the spiral field generated at the halo/WIM interface. We have a suitable energy source in the shear itself. We expect this to dissipate field energy through diffusivity and reconnection -- here see \cite{Wezgowiec:2016aa} who discuss reconnection as a possible source for the heating of the hot gas above the magnetic interarm regions in NGC 6946.\footnote{The importance of field and energy generation at such a disc/corona interface is also discussed by \citet[Section 7.2.4]{Mayer:2007aa} in the context of black hole accretion discs.}

 We also note that \cite{Heald:2016aa} discuss the possibility of a large-scale regular magnetic field in the disc--halo interface of M83, and \cite{Mulcahy:2017aa} discuss the halo/WIM shear region in NGC 628 as a possible source of magnetic turbulence.

\subsection{Magnetic field and density setup}
\label{fieldstruct}
In this subsection we describe the magnetic field structure of our toy model. Following the arguments above, we take the WIM to have a height of $500$\,pc, a density of $n = 0.3$\,cm$^{-3}$ and a temperature of $10^4$\,K. We also take the halo to have a height of $5$\,kpc, a density of $0.003$\,cm$^{-3}$ and a temperature of $10^6$\,K.

For the magnetic field structure we follow \cite{Henriksen:2016aa} and \cite{Lubow:1994aa}. From \cite{Henriksen:2016aa} we take the azimuthal velocity to be
\begin{equation}
  v_\phi = \frac{V}{R/R_0}\left(\sqrt{\left(R/R_0\right)^2 + \left(z/z_0\right)^2}-z/z_0\right)\,,
\end{equation}
where we take $R_0 = 5$\,kpc and $z_0 = 500$\,pc as our scalelengths. From this velocity field we construct the azimuthal magnetic field, $B_\phi$, proportional to the azimuthal velocity, as
\begin{equation}
  \label{Bphi}
  B_\phi = -\frac{B_{\phi,0}}{R/R_0}f(z)\left(\sqrt{\left(R/R_0\right)^2 + \left(z/z_0\right)^2}-z/z_0\right)\,,
\end{equation}
where $B_{\phi,0}$ is a free parameter determining how bent the magnetic field lines are in response to the vertical shear. The minus sign ensures that the field exhibits trailing spirals as required. We have also included an additional function $f(z) = z_0/(z+z_0)$ to smoothly remove the azimuthal component of the field at large distances from the galaxy plane. We note that (\ref{Bphi}) is independent of $\phi$ and thus the $\phi$ component of the divergence constraint is zero. \note{We also note that for the parameters we have taken the WIM/halo scaleheights are independent of radius, compared with real galaxies which are flared -- we do not anticipate this affecting our results.} 

\begin{figure}
  \includegraphics[width=\columnwidth]{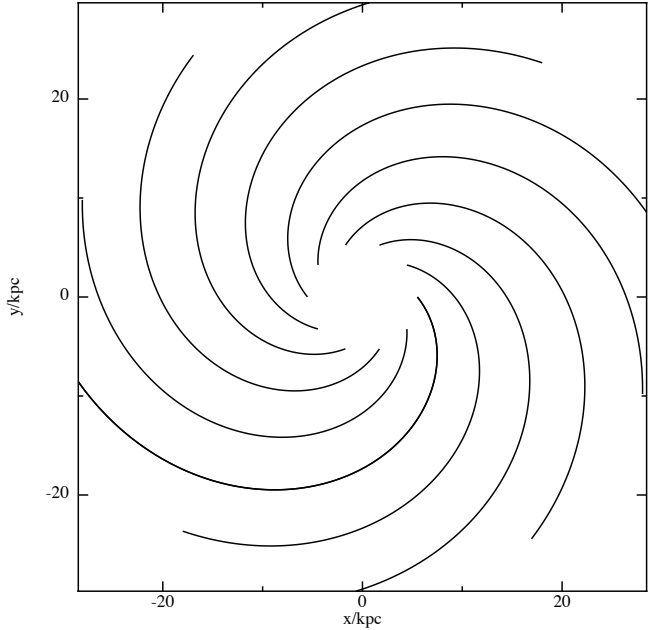}
  \caption{Magnetic field lines projected on to the $x$-$y$ (galaxy) plane. Ten field lines are shown, each originating on the $z=0$ plane at $R=5.5$\,kpc and extending until $z = 5$\,kpc. The field lines show trailing spirals as the galaxy rotation is counter-clockwise. This occurs due to angular momentum transport, and hence radial flow, created by the vertical shear and mediated by the magnetic field. The halo gas has lower angular momentum than the WIM, and is therefore spun outwards.}
  \label{fig1}
\end{figure}

\begin{figure}
  \includegraphics[width=\columnwidth]{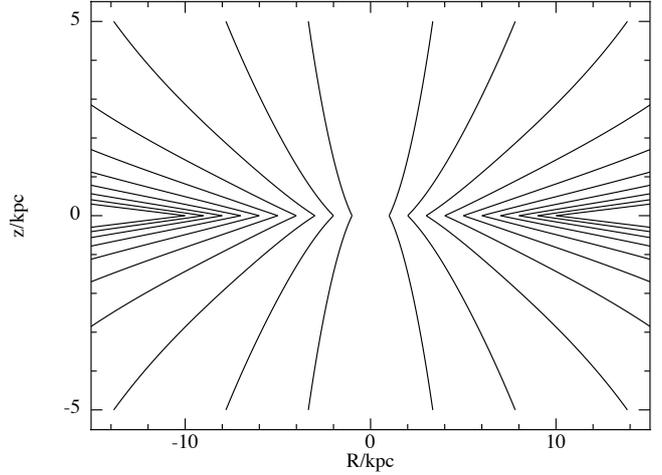}
  \caption{Magnetic field lines in the $R$-$z$ plane. These lines are created assuming $B_\phi$ is zero. If this plot were created in 3D, with $B_\phi$ included, the field lines would also bend in azimuth as depicted in Fig.~\ref{fig1}. This figure shows that the ratio $B_z/B_R$ is not far from unity for most of the WIM/halo. At small $|z| < 100$\,pc our model is not relevant as we neglect the dense galactic plane where e.g. star formation is taking place. The field appears pinched at small $z$ as an initially vertical field is spun outwards at large $z$ due to angular momentum transport mediated by the magnetic field.}
  \label{fig2}
\end{figure}

For the radial and vertical components of the field we aim to match the findings of \cite{Lubow:1994aa} who also considered the consequence of dragging field lines in discs. We thus create a field function for the $R$-$z$ components as
\begin{equation}
  \psi = \psi_\infty + \psi_{\rm disc} = \frac{1}{2}B_0 R + \frac{1}{2}B_1\frac{z_0^2}{z + z_0}\tanh(R/R_0)\,.
\end{equation}
We therefore have
\begin{equation}
  \label{Br}
  B_R = -\frac{\partial\psi}{\partial z} = \frac{1}{2}B_1\frac{z_0^2}{(z + z_0)^2}\tanh(R/R_0)
\end{equation}
and
\begin{equation}
  \label{Bz}
  B_z = \frac{1}{R}\frac{\partial}{\partial R}(R\psi) = B_0 + \frac{1}{2}B_1\frac{z_0^2}{z + z_0}\left(\frac{\tanh(R/R_0)}{R} + \frac{{\rm sech}^2(R/R_0)}{R_0}\right)\,.
\end{equation}
Inspection of these equations suggests that we take $B_{\phi,0} = B_1/2$ and $B_0 = 0.1B_1 z_0/R_0$ to achieve the field structure we expect. For clarity we plot these fields in Figs.~\ref{fig1} \& \ref{fig2}. Along the field lines, the field strength varies by a factor of a few, so we normalise the field strength to a canonical value $B_{\rm c}$ at $R=1\,$kpc and $z=0.1\,$kpc, and it is this value we will quote in the discussion below.

For the polarization calculations in the next section we take an inner radius $r_{\rm in} = 1$\,kpc and an outer radius of $r_{\rm out} = 10$\,kpc. For the vertical extent we employ $z_{\rm in} = 0$ and $z_{\rm out} = 5$\,kpc. We note that the galaxy midplane is expected to be Faraday thick, so we do not expect any contribution from emission below the plane $z<0$.

To find the total polarized flux from this simple model, we must make some estimate of the emissivity of polarized flux throughout the galaxy, and then calculate the Faraday depolarization along the line-of-sight to the observer. We describe this calculation and the results in the next section.

\section{The polarized emissivity, flux and Faraday depolarization}
\label{sec:polarflux}

In Cartesian coordinates $ (x,\,y,\,z)$, the galaxy rotates around the $z$-axis. From Section~\ref{fieldstruct} above we have an estimate of the field ${\bf B}({\bf x})$ and number density $n({\bf x})$. If we suppose that an observer lies in the direction $(0,\sin i,\cos i)$, then seen from the observer $Ox$ is the galactic major axis, and $Oy$ is the galactic minor axis. Using this we can define the set of coordinate $(x^{\prime},\,y^{\prime},\,z^{\prime})$, where $z^\prime$ is parallel to the line-of-sight of the observer. In these new coordinates the magnetic field components are given by
\begin{equation}
  B^{\prime} = (B_x^{\prime},B_y^{\prime},B_z^{\prime}) = (B_x,B_y\cos i - B_z\sin i,B_y\sin i + B_z\cos i)\,.
\end{equation}

In the primed coordinate system, we must now integrate along the line-of-sight through the galaxy to the observer, calculating the emissivity and Faraday depolarization to make maps of the total polarized flux seen by the observer. We thus define each midplane point ${\bf x}_0 = (x_0,y_0,0)$, and compute the integrals along the path ${\bf x}(s) = (x_0,y_0+s\sin i,s\cos i)$ with $s=0$ being the galactic plane.

To estimate the polarized emission seen by the observer when looking towards this point we need to know (i) the polarized emissivity and polarization angle at each point along the line of sight and (ii) the Faraday rotation between that point and the observer. Cosmic ray electrons (CRE) are assumed to arise mostly from supernovae, and these are predominantly confined to the galactic plane. However, the CRE propagate through the galaxy. \cite{Sun:2008aa} discuss a model where the CRE density is an exponential function of vertical height with scalelength $\sim 1$\,kpc. Therefore the CRE density falls off with vertical height not too differently to the gas (thermal) density we have assumed. Thus, for simplicity, we assume that the CRE density $n_{\rm e}$ is equal to the gas number density $n$. Assuming the CRE density is constant with radius appears reasonable \citep{Strong:1996aa}. For simplicity, as we do not know the cosmic ray energy density, we take the polarized emissivity between $s$ and $s+{\rm d}s$ as
\begin{equation}
  {\rm d}I_p \propto n_e B_\perp^2{\rm d}s\,,
\end{equation}
where $B_\perp^2 = {B^\prime_x}^2 + {B^\prime_y}^2$ is the field strength perpendicular to the line-of-sight, and therefore controls the polarized emissivity in the direction of the observer from synchrotron. Then the total emitted polarized flux along a line of sight is
\begin{equation}
I_p(x_0,y_0) = \int_{s=-\infty}^{s=\infty}{\rm d}I_p\,,
\end{equation}
assuming that the galaxy is optically thin to radio emission at the relevant wavelength. We assume in this calculation a wavelength of 20\,cm.

However, the emitted polarized flux can add together along the line-of-sight to depolarize. This occurs if the field direction changes along the line of sight, and also through Faraday rotation. The local polarization angle of the emitted flux is given by
\begin{equation}
  \chi_0 = \tan^{-1}\left(\frac{B_y^\prime}{B_x^\prime}\right)\,.
\end{equation}
Then the final polarization angle of the emission reaching the observer is given by $\chi = \chi_0 + \Phi(s)$, where $\Phi(s)$ is the total Faraday rotation along the line-of-sight from $s$ to the observer, given by (\ref{fara}) above. From this we can calculate the total polarized flux as $\left|{\bf P}\right|$, where ${\bf P}$ is the vector including the polarization direction. ${\bf P}$ is defined as
\begin{equation}
  {\bf P}(x_0,y_0) = (Q,U) = \int_{s=-\infty}^{s=\infty}{\rm d}I_p (\cos2\chi,\sin2\chi)\,,
\end{equation}
where the $2\chi$ comes from considering Stokes' vector addition, and the final polarization direction for the line-of-sight $(x_0,y_0)$ is given by the angle $\chi_f = 0.5\tan^{-1}\left(U/Q\right)$.

In this calculation we have assumed that the emissivity roughly follows the perpendicular line-of-sight field strength (squared, $B_\perp^2$), which follows from synchrotron. This of course differs from the Faraday rotation, which instead follows the parallel line-of-sight field strength ($B_\parallel$). Below we will show results for wavelengths $\lambda = 6$, $20$\,cm and magnetic field strengths $B_{\rm c} = 1$, $4\,\mu$G. This yields ${\rm FD(halo)} \approx 15B_{\rm c}$ and ${\rm FD(WIM)} \approx 150B_{\rm c}$. Recalling that at $6$\,cm ${\rm FD}_{\rm cr} = 280$ and at $20$\,cm ${\rm FD}_{\rm cr} = 25$, we can see that for $B_{\rm c}=1\,\mu$G, we expect to see through most of the WIM at $6$\,cm, but only just past the halo at $20$\,cm. Whereas for $B_{\rm c}=4\,\mu$G, we can see partially into the WIM at $6$\,cm, and not even through the halo for $20$\,cm. Therefore, for these parameters, the observable polarization comes from the WIM/halo, but not from deeper into the galaxy plane. We note that for magnetic field strength of $B_{\rm c}=4\,\mu$G, the field is in equipartition with the plasma $\beta \sim 1$ (see eq~\ref{equiB} above).

\begin{figure}
  \includegraphics[width=\columnwidth]{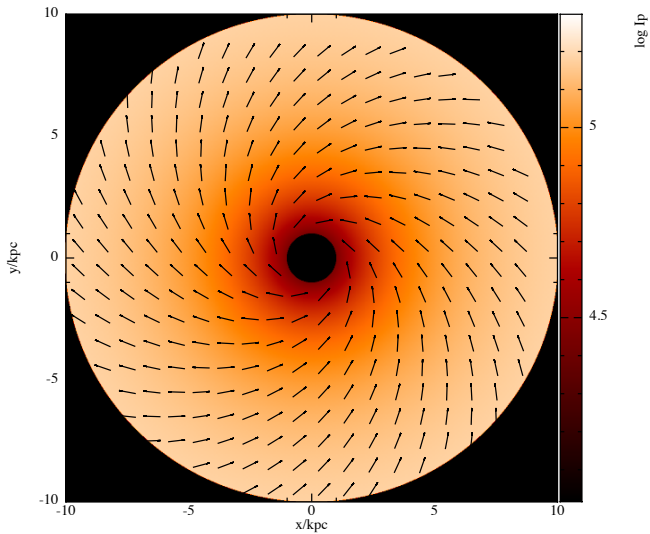}
  \caption{A mock representation of a galaxy viewed face-on. The colour represents the emissivity of polarized flux. The black lines represent the polarization B-vectors. In this case the field has a field strength of $B_{\rm c} = 1\,\mu$G, and the assumed wavelength is $6$\,cm. In this case there is only small Faraday rotation of most of the emission from the WIM, and thus the polarization tracks the field direction in this region. This structure is similar to those observed for e.g. NGC 4736 \citep{Beck:2013aa}.}
  \label{fig3}
\end{figure}

\begin{figure}
  \includegraphics[width=\columnwidth]{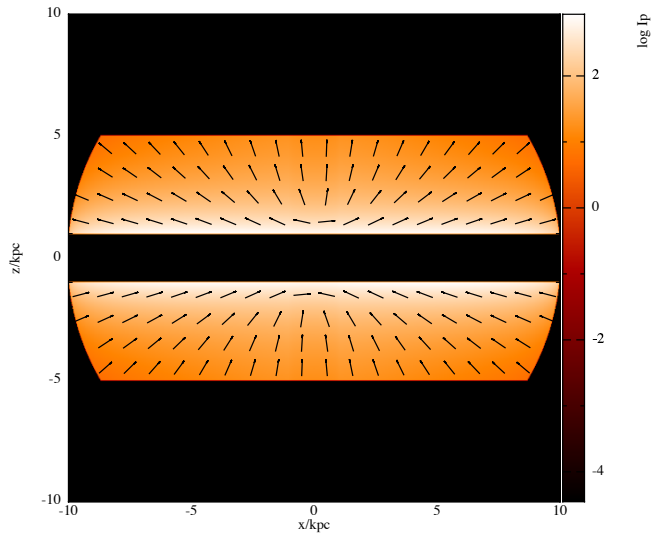}
  \caption{The same as Fig.~\ref{fig3} but viewed edge-on. The colour represents the emissivity of polarized flux. The black lines represent the polarization B-vectors. In this case the field has a strength of $B_{\rm c} = 1\,\mu$G, and the assumed wavelength is $6$\,cm. We have blanked out the emission from the midplane to reinforce that our model is not relevant for emission from the galactic plane. The field lines show a larger radial component at small $z\sim z_{\rm WIM}$ and a larger vertical component at $z \sim z_{\rm halo}$. This structure is similar to those observed for e.g. NGC 891 and NGC 4631 \citep{Krause:2009aa}.}
  \label{fig4}
\end{figure}

\begin{figure*}
  \includegraphics[width=\textwidth]{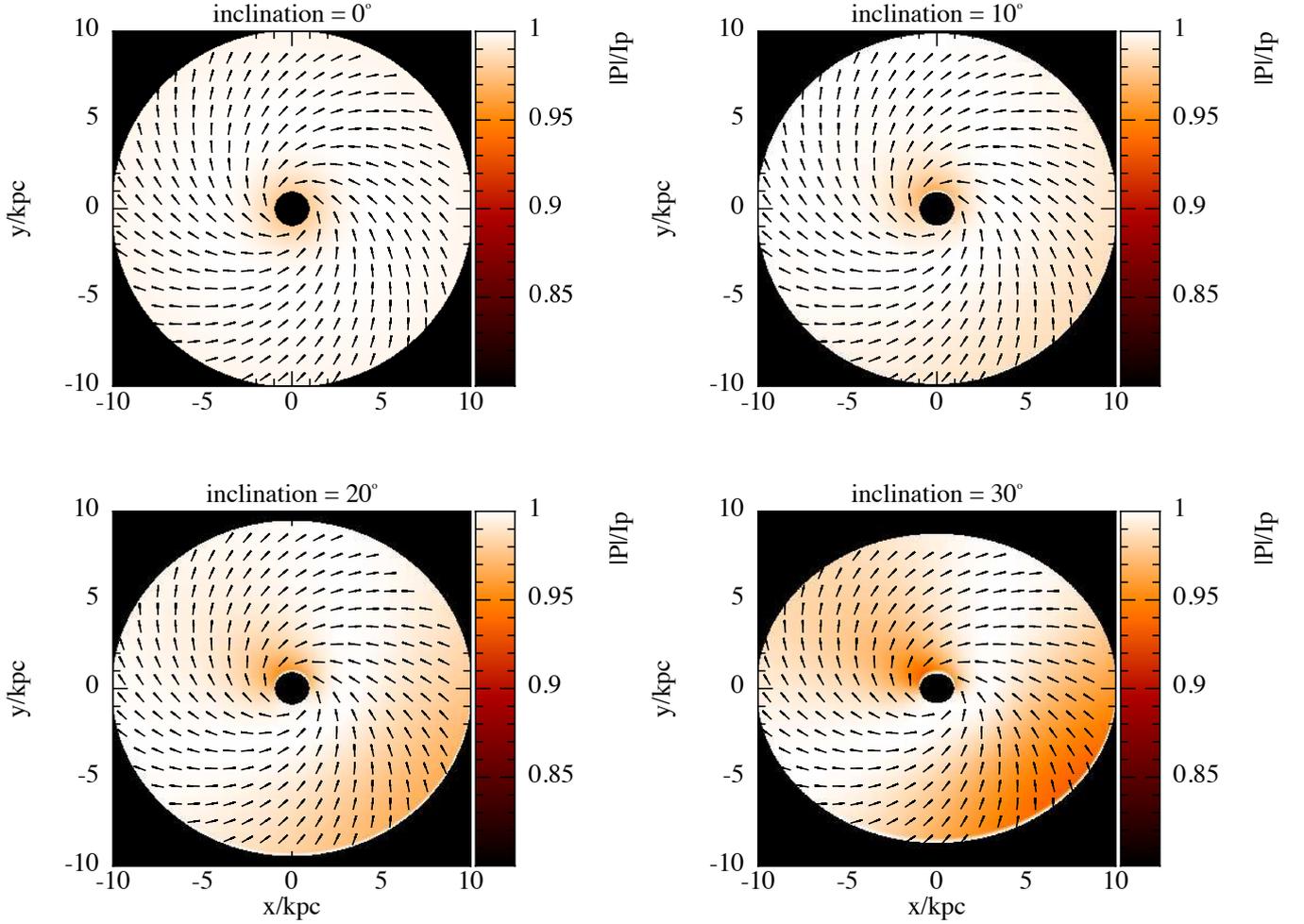}
  \caption{The fractional polarized flux, with polarization B-vectors, for four different inclinations. In this case the field has a strength of $B_{\rm c} = 1\,\mu$G, and the wavelength is $6$\,cm. The plots show the observed polarized flux in this case is 90-100 per cent of the total emitted polarized flux. This is as expected for these parameters. As the inclination is increased an asymmetry appears, with less flux emitted on the receding side (recall that the galaxy rotates counter-clockwise, \note{and as the inclination angle is increased the top half of the galaxy moves into the page}), but the change in polarized flux is minimal.}
  \label{fig5}
\end{figure*}

\begin{figure*}
  \includegraphics[width=\textwidth]{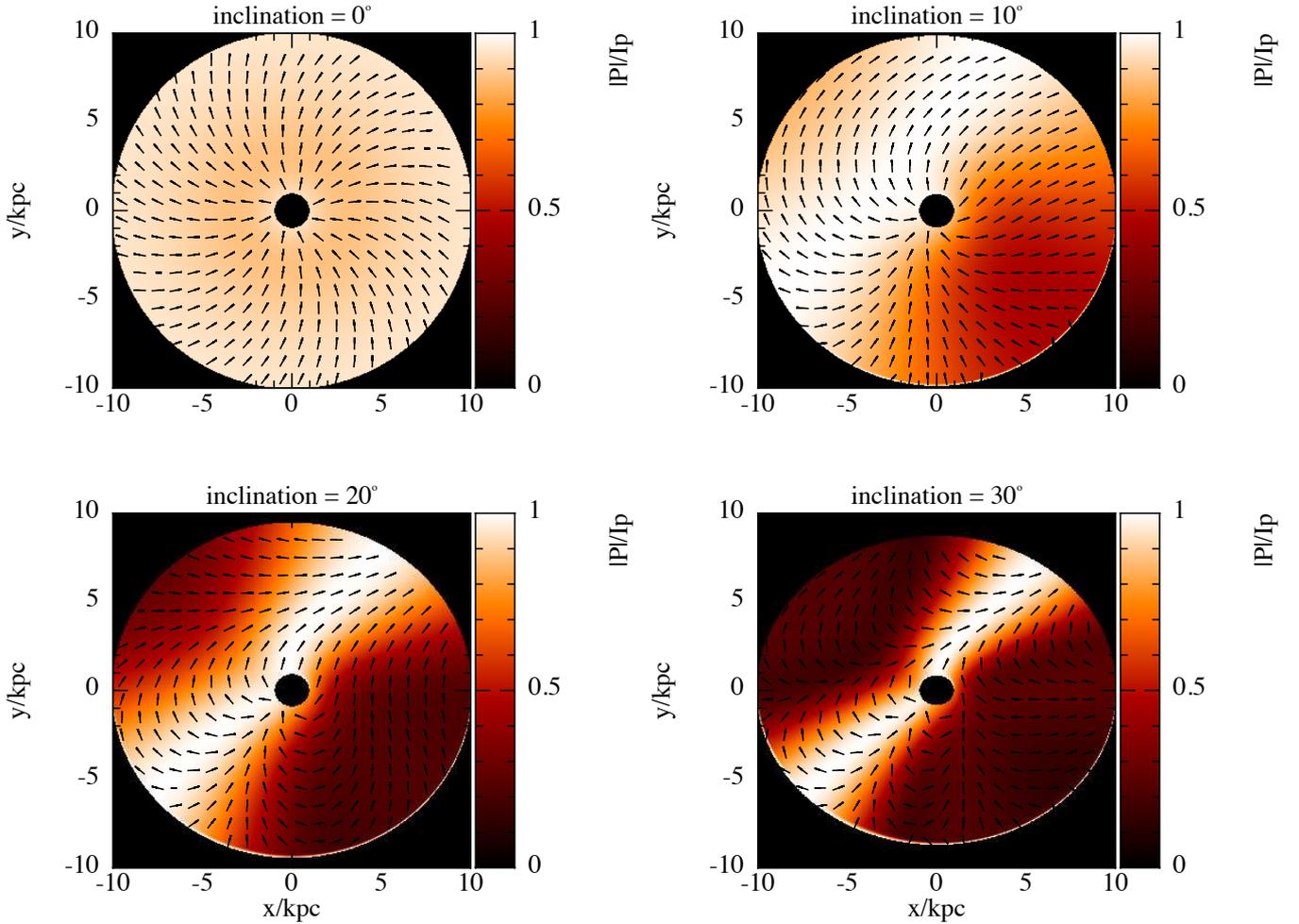}
  \caption{The same as Fig.~\ref{fig5} but for a wavelength of $20$\,cm. In this case the polarized flux is reduced by Faraday depolarization. Even in the face on case there is a small reduction in flux. As the inclination is increased there is a dramatic drop in flux on the receding side of the galaxy (recall that the galaxy rotates counter-clockwise). At $20-30^\circ$ inclinations there is also a drop in flux on the approaching side. The resulting pattern of polarized intensity with azimuth is reminiscent of some of the observed data presented in Fig.~1 of \protect\cite{Braun:2010aa}. In this case the regions of lower polarized flux contains ordered, but rotated, B-vectors -- indicating that the Faraday rotation angle is order unity, and not several rotations (cf. Fig.~\ref{fig8} below). \note{Note that in this figure the colour bar scale is larger than in Fig.~\ref{fig5} above -- in Fig.~\ref{fig5} the colour scale was condensed to highlight small variation, while here the variations are substantial.}}
  \label{fig6}
\end{figure*}

\begin{figure}
  \includegraphics[width=\columnwidth]{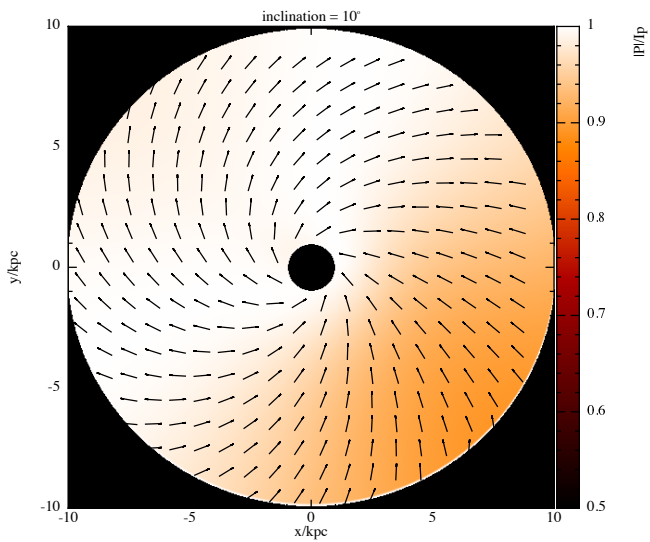}
  \caption{This figure is the same as the $10^\circ$ case in Fig.~\ref{fig5} (i.e. a wavelength of $6$\,cm), but here with a field strength of $B_{\rm c} = 4\,\mu$G. This results in stronger Faraday rotation. Thus there is a larger drop in polarized flux on the receding side of the galaxy. At significantly larger field strengths the emission would be strongly depolarized, which at this wavelength and inclination is generally not seen.}
  \label{fig7}
\end{figure}

\begin{figure}
  \includegraphics[width=\columnwidth]{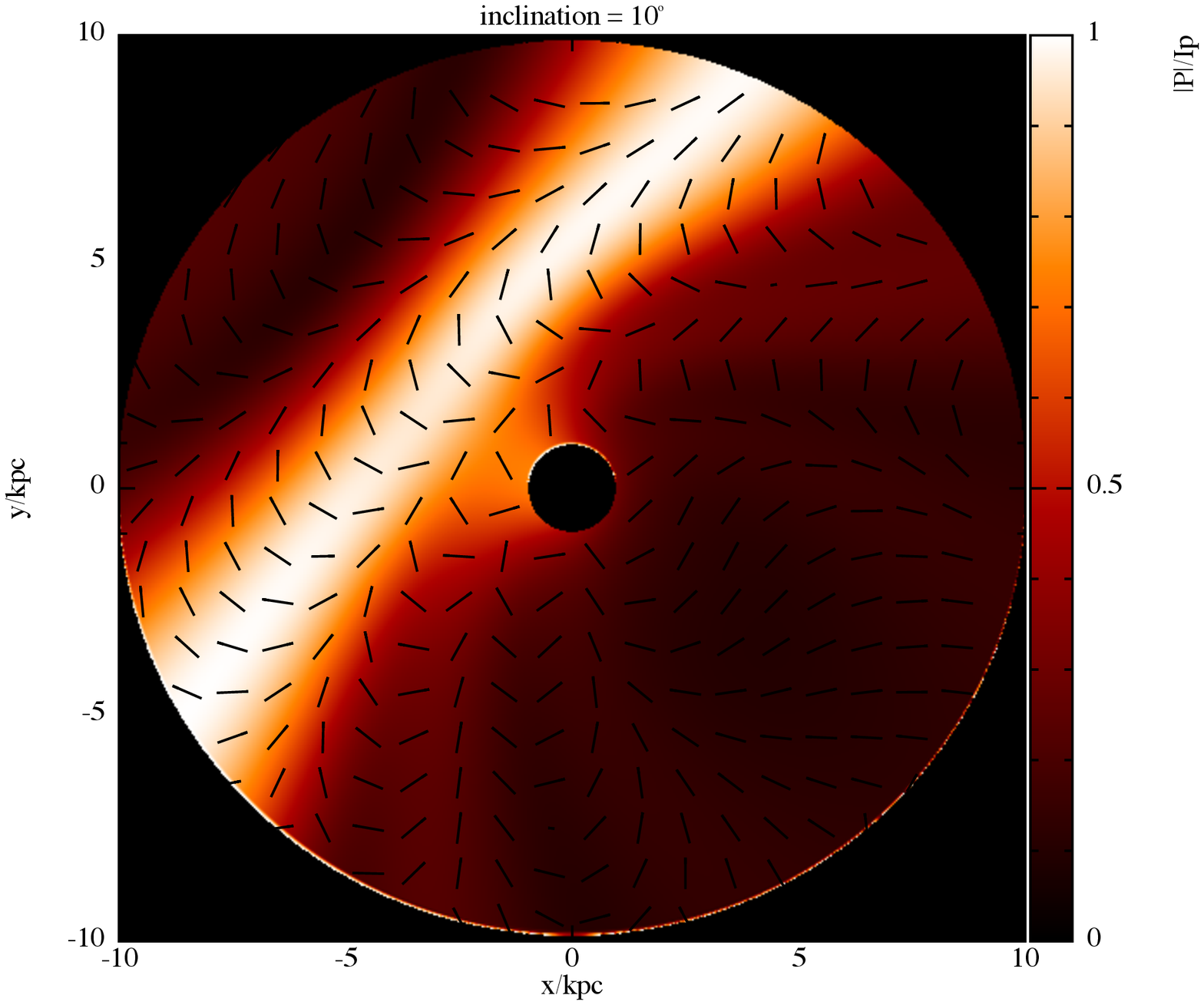}
  \includegraphics[width=\columnwidth]{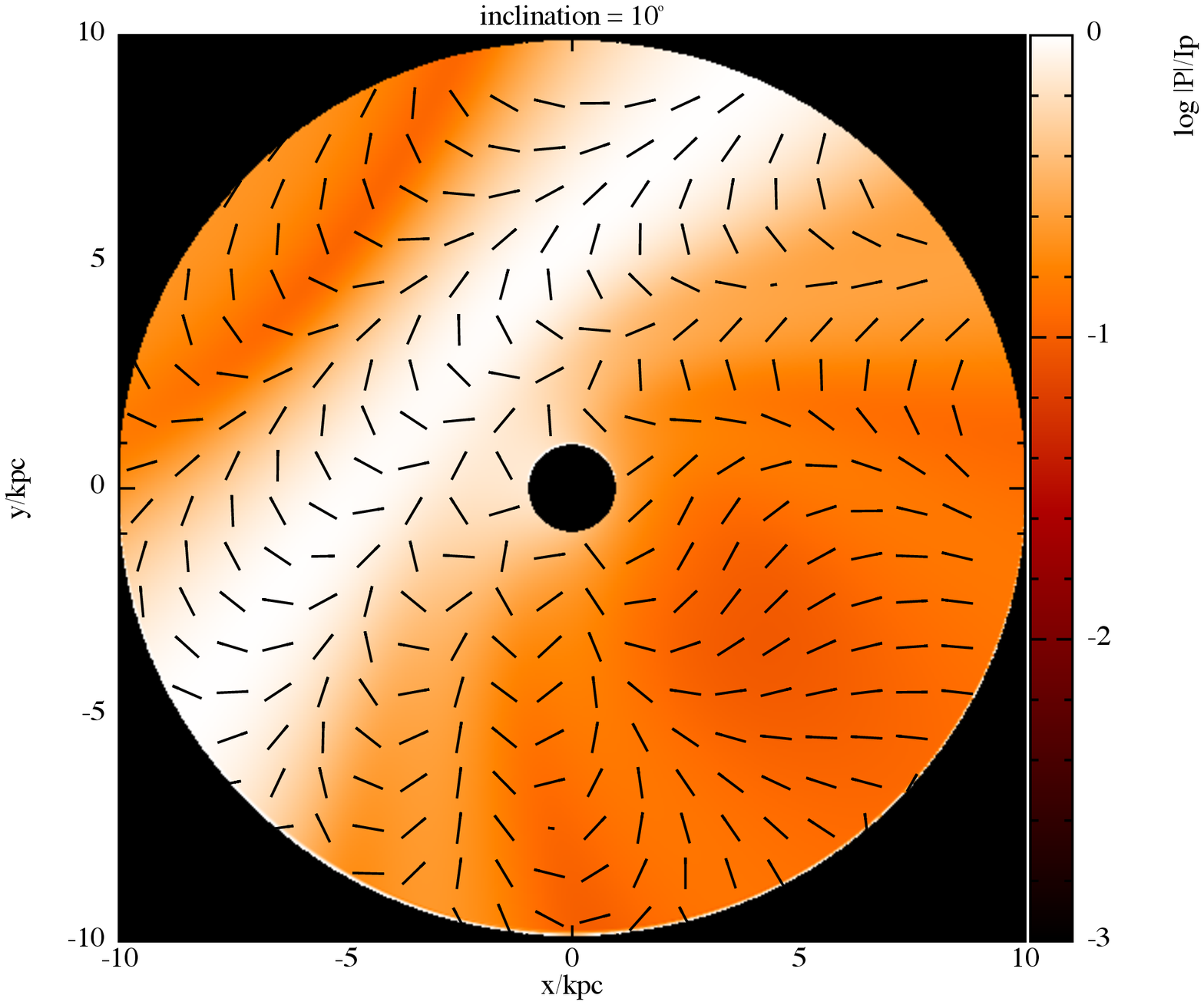}
  \caption{Top panel: This figure is the same as the $10^\circ$ case in Fig.~\ref{fig6} (i.e. a wavelength of $20$\,cm), but here with a field strength of $B_{\rm c} = 4\,\mu$G. This results in very strong Faraday rotation. Thus there is almost no polarized flux on the receding side of the galaxy, while some remains on the approaching side. The polarization vectors are strongly rotated with no obvious pattern in the low polarized flux regions. Bottom panel: Same as top panel but with polarized flux shown on log scale to highlight the polarization vectors in regions of low flux.}
  \label{fig8}
\end{figure}

Notwithstanding the caveats and assumptions we have discussed throughout, our results are encouraging. We show in Figs~\ref{fig3} \& \ref{fig4} the emissivity and polarization B-vectors for the face-on and edge-on cases where the field strength is $B_{\rm c}=1\,\mu$G and the wavelength is $6\,$cm. In these cases there is only small amounts of Faraday rotation, and the polarization B-vectors track the magnetic field structure. In the face-on view the polarization B-vectors show a trailing spiral as observed, and in the edge-on view they show the observed X-shaped morphology. To explore the effects of inclination on the polarized flux, we plot the fractional polarized emission for several different inclination angles in Fig.~\ref{fig5} using the same parameters. At higher inclinations ($\sim 30^\circ$) there is a very modest drop in polarized flux, and the spiral pattern is preserved. A contrasting picture is presented in Fig.~\ref{fig6}, where the same inclinations are shown, but this time at $20\,$cm wavelength. In this case the polarized flux is strongly diminished by Faraday depolarization, and for larger inclinations the simple spiral polarization pattern is altered. The most important result is that the polarized flux is predominantly reduced on the receding side of the galaxy for even modest inclination angles. For angles $\sim 20-30^\circ$ the polarized flux is essentially lost on the receding side, and the emission varies with azimuth in a similar manner to some of the observed galaxies in Fig.~1 of \cite{Braun:2010aa} with a double peak. Finally for comparison we show in Fig.~\ref{fig7} the $10^\circ$ inclination case with a magnetic field strength of $B_{\rm c} = 4\,\mu$G at a wavelength of $6\,$cm, and the same in Fig~\ref{fig8} at $20\,$cm. As predicted these show much larger variations due to Faraday rotation. Fig.~\ref{fig7} shows a modest drop in polarized flux on the receding side, whereas Fig.~\ref{fig8} shows no polarized flux on the receding side, and little polarized flux on the approaching side. In this case, at $20\,$cm, the polarization B-vectors show no obvious pattern as expected.

\subsection{Summary}
\note{These results suggest that this simple model is capable of reproducing the three well-known \citep[e.g.][]{Beck:2016aa} observed features of large-scale galactic magnetic fields discussed in the Introduction, namely:}
\begin{enumerate}

\item \note{Galaxies seen edge-on display X-shaped fields. This is caused by the pinching of an initial global 
vertical field by the radial inward/outward motion of the WIM gas/halo gas, respectively,}

\item \note{Galaxies seen face-on show spiral magnetic patterns. These are caused by the twisting of the initial global vertical field by the rotation of the galactic disk (WIM), and}

\item \note{In galaxies seen at intermediate angles, and at longer wavelengths ($\lambda \sim 20$ cm) we observe a reduction in the polarized emission on the receding end of the (apparent) major axis. This comes about because the line of sight towards the receding end of the major axis tends to lie along the (twisted and pinched) field lines. This means that, compared to the approaching end of the major axis, the emission is less polarized, and that Faraday rotation is enhanced, especially at longer wavelengths.}
\end{enumerate}

\section{Spiral structure}
\label{sec:spiral}
We have shown that the three characteristics of observed large--scale galaxy magnetic fields can be reproduced by a simple picture in which an initial poloidal field is sheared by gas motions, without invoking dynamo action. \cite{Henriksen:2016aa} reach a similar conclusion. As we noted in the Introduction, dynamo theory is also invoked to explain the affect of galaxy spiral structure on the galaxy magnetic fields. Here we consider whether the simple model we have described can do this.

\subsection{Observations and current interpretation}
Observations of polarization structures at wavelengths $\lambda \sim 3 - 6$\,cm in spiral and disc galaxies are assumed to be essentially free of Faraday rotation. The polarization structures are interpreted in terms of the presence of ordered fields in the galaxy disc planes. Spiral patterns are found in almost every galaxy, even those lacking in optical spiral structure, like the ringed galaxy NGC 4736, which seems to display an axisymmetric spiral pattern. \cite{Beck:2013aa} comment that if large--scale fields were frozen into the gas, the galactic differential rotation would wind them up to very small pitch angles. They argue that the observed pitch angles of 10 -- 40 degrees indicate a general decoupling between the magnetic fields and the gas flow due to (effective) magnetic diffusivity, and claim that this is a strong indication for $\alpha - \Omega$ dynamo action. However we have discussed above an idealized model (Sections~\ref{sec:simple} \& \ref{sec:polarflux}) in which the winding--up timescale for the field is not the galactic shear timescale, which corresponds to a galactic rotation timescale of  $\sim$ 300 Myr, as assumed by \cite{Beck:2013aa}, but is instead the timescale for the shear between the WIM and the halo \citep[see also][]{Henriksen:2016aa}. This timescale is longer than the galactic rotation timescale by a factor of $\sim \Omega/ \Delta \Omega \sim 10$ (Section 5), even without allowing for any magnetic diffusivity.

In most spiral galaxies the highest polarized intensities (i.e. the strongest ordered fields) are detected {\em between} the optical arms, filling a large fraction of the interarm space \citep{Beck:2016aa}, and the magnetic pitch angles differ systematically from those of optical spiral structures. This is sometimes given as evidence for an $\alpha-\Omega$ dynamo \citep{Beck:2016aa}, although we note that this type of model can produce a wide range of pitch angles by suitable adjustment of poorly constrained parameters, so that this is not a strong test. Indeed \cite{Chamandy:2015ab} point out that observed magnetic pitch angles are systematically larger than predicted by standard mean field dynamo theory. They note that the observed values can be obtained by dynamo parameters located towards the boundary of allowed parameter space, and suggest that additional physical effects may need to be incorporated. \note{However, \cite{Chamandy:2016aa} find agreement between observed values and pitch angles predicted from standard mean field dynamo theory, but they caution that the resulting field strengths and growth rates are strongly dependent on unconstrained parameters.} It is also sometimes argued \citep[e.g.][]{Mulcahy:2017aa} that the smoothness of the magnetic pitch angle is similarly consistent with dynamo action because the large--scale field is built up over several galactic rotations and is not strongly affected by local features in the arms. However, this argument would imply that dynamo action is too slow to reinstate ordered fields downstream of the arms.

\subsection{The nature of spiral arms}
Before considering the relationship between global magnetic structures and (optical) spiral arms, it is important to understand the nature of the spiral arms themselves. The simple ideas that spiral arms are long--lived self--reinforcing structures, viewed as dynamical modes of the underlying single disc, and therefore with a single pattern speed $\Omega_P$ (Lin--Shu density waves, \citealt{Lin:1964aa}) is no longer tenable. Nevertheless much, though not all,  of the work on the dynamo generation of magnetic arms \citep[see, for example,][]{Chamandy:2013aa,Chamandy:2013ab,Moss:2013aa,Moss:2015aa} makes this assumption. One benefit of this assumption is that it allows a time comparable to the age of the galaxy ($\sim 10$ Gyr, \citealt{Chamandy:2015aa}; $\sim 8 - 13$ Gyr, \citealt{Moss:2015aa}) for the build up of dynamo generated spiral features. The current view \citep{Dobbs:2014aa} is that instead spiral arms are dynamically evolving quantities, with no overall pattern speed, and with lifetimes comparable to galactic shear timescales of $\sim$ 0.1 -- 0.2 Gyr, much less than the age of the galaxy. They are more akin to `kinematic density waves' \citep{Toomre:1972aa,Binney:1987aa}, or more accurately, density patterns,  which rotate locally with local angular velocity $\Omega_P \approx \Omega - \kappa/2$. Since $\Omega_P$ is a function of radius, the pattern is subject to shear. This implies that spiral arms must be excited and/or continuously regenerated \citep{Toomre:1972aa}. In barred spirals, which we consider only briefly here, the arms are often excited by the bar, at least in the inner regions. Apart from this, to a first approximation, there are two mechanisms by which spiral arm (re--)generation is thought to occur.

\subsubsection{Tidal interactions -- grand design spirals}
\label{sec:tidal}
The spiral arms in galaxies such as M51 and M81 are clearly due to tidal interaction. \cite{Dobbs:2010ab} present a dynamical model of the tidal interaction of M51 (NGC 5194)  with its neighbour NGC 5195. They assume that the disk gas has a fixed temperature of $10^4$ K. The interaction has been occurring over the last $\sim 200$ Myr and the arms originate primarily from two close passages between the two galaxies.  The  two sets of arms thus generated evolve and merge over that timescale to give the underlying current grand design spiral. The rapidly evolving structure (the arms wind up on a timescale of $\sim 100$ Myr, see also \citealt{Meidt:2009aa}) also produces the asymmetry between the arms, as well as short-lived kinks and bifurcations. The flow of gas through such rapidly evolving arms does not follow the traditional expectations of Lin--Shu density wave theory in that gas flow across the arms, as delineated by the star--formation history, is not steady and monotonic \citep{Dobbs:2010aa,Foyle:2011aa}.

The spiral arms in M81 (whose magnetic properties are presented by \citealt{Sokolov:1992aa}) also arise from tidal interactions. Modelling of the interaction \citep{Yun:1999aa} indicates that tidal structures were induced by the passages of the two companions M82 around 200 Myr ago and NGC 2077 around 280 Myr ago. As before, the interactions between the two sets of spiral arms thus induced produce the observed bifurcations and kinks in the current arms. Here again the evidence for the lack of monotonic star formation chronology across the arms \citep{Choi:2015aa} suggests that the traditional idea of Lin--Shu density wave theory of a steady spiral pattern with gas flowing steadily through the arms at each radius is not tenable. \note{This also implies that the concept of a co-rotation radius, of crucial importance to many spiral dynamo models (see below) is invalid.}

In both these galaxies the spirals arms have been caused by tidal interactions which occurred only one or  two galactic rotation periods in the past. This in turn implies that interpreting any magnetic structures in these galaxies in terms of the long--term modal structure of standard dynamo theory is problematic. 

\subsubsection{Local self--gravity -- flocculent spirals}
In the absence of forcing (for example tides from external interactions or an internal bar) it is still possible for spiral structure to form provided that the stellar disc is sufficiently self--gravitating. This draws on the idea of swing amplification \citep{Toomre:1981aa}. A number of recent simulations \citep{Fujii:2011aa,Wada:2011aa,Baba:2013aa,Baba:2015aa,DOnghia:2013aa,DOnghia:2015aa} demonstrate that local self--gravity acts as a seed to form a local segment of a spiral arm by swing amplification. The rotational speeds of such spiral arms follow the local rotation speed  (i.e. each part of the arm can be regarded as a corotation point).The arms are non--steady. They are wound and stretched by the galactic shear, and thus the local properties of the arm change on a timescale of around 100 Myr, as they bifurcate and/or merge with other arms. In such galaxies the merging of arms can give apparent arm structures extending over a large range of radii. The number of arms depends on the galaxy properties at each radius, so that the number of arms can change with radius \citep{Fujii:2011aa,DOnghia:2015aa}. \cite{Baba:2015aa} suggests that it is this mechanism which creates the grand--design or large--scale spiral arms in barred spiral galaxies. A possible example here could be M83.

Thus, for example, the galaxy IC342, which shows a fairly uneven pattern of spiral arms, with two arms in the inner disk and four arms in the outer disk \citep{Meidt:2009aa}, could well fit this picture. The disc in NGC 6946 shows an irregular spiral structure with flocculent and bifurcating arms. The arm dynamics for this galaxy are complicated since the disc is strongly self--gravitating \citep{Romeo:2015aa}, but the galaxy also has close companions \citep{Pisano:2014aa}.

In these cases the local arms are shortlived and essentially corotating with the gas, so that any motion of gas through the arms is minimal \citep{Dobbs:2010aa,Wada:2011aa}. This implies that interpreting the magnetic structures in such galaxies in terms of the interaction between local dynamos and Lin--Shu style flow of gas through density waves or in terms of combinations of longlived dynamo modes is not likely to give an accurate representation of reality.

\subsection{Spiral patterns in galactic magnetic fields}
We have seen that the typical lifetimes for spiral arm features in disk galaxies are now thought to be on the order of the shear timescale (or a galactic rotation time) of $100 - 200$\, Myr. This, by construction, is also the growth timescale for local galactic dynamos (although, as we have argued above, the timescale for the generation of a {\it large--scale} field by such dynamo action is probably much longer). This implies that interpreting magnetic field structure in terms of local dynamo action has problems with timescales.

In addition it is important to realise that the optical spiral arms are visible because of the presence of dense (molecular) clouds and young stars which form within them. Within a spiral shock, these clouds form from material that is not typical of the ISM, but which is already cooler and denser \citep{Pringle:2001aa,Dobbs:2012aa}. In addition, because in a typical grand--design spiral the shock velocity is around $10 - 20\,{\rm km\,s^{-1}}$, the WIM (at temperatures of $\sim 10^4$ K) only undergoes weak shocks, with Mach number at most ${\cal M} \sim 1 - 2$. Rather it is the cooler elements of the ISM ($T \sim 100$K) which shock strongly enough to become self--gravitating and so form stars \citep{Dobbs:2006aa,Wada:2011aa}. In flocculent spiral galaxies, the spiral arm segments corotate locally and thus the gas flow relative to the spiral arms is smaller by about an order of magnitude \citep{Wada:2011aa}. This implies that in these galaxies, the WIM (which is supposedly the seat of galactic dynamo activity) does not partake in any shocks, and indeed barely reacts at all. The exception to this is in a case such as M51  where the tidal interaction is sufficiently strong (indeed NGC 5194 and NGC 5195 are predicted to merge in the near future) that even gas at temperatures of $10^4$K is able to shock \citep{Dobbs:2010ab}.

\subsubsection{\note{Spiral patterns in dynamo theory}}
\note{As we have noted, the dynamo theory models are currently too crude to distinguish between the cool, mainly molecular, gas which lies close to the plane $|z| \le 100$ pc where spiral shocks, star formation and resultant energy injection by hot stars and supernovae occur, and the WIM, where dynamo activity is modelled (in the thin disk approximation). \cite{Chamandy:2013aa} and \cite{Moss:2013aa} model spiral magnetic arms in galactic dynamos by imposing non-axisymmetric forcing of the mean-field dynamo in a spiral pattern. These models use a rigidly rotating spiral, and produce non-axisymmetric fields predominantly near the co-rotation radius. As we have remarked above (Section~\ref{sec:tidal}) this is not what is seen in the grand design spirals. They do, however, demonstrate that the time taken to produce a coherent large-scale field across the whole galactic disk is, as expected, comparable to the age of the galaxy (see Section~\ref{sec:timescales}).}

\note{\cite{Chamandy:2015aa} generalise the concept of non-axisymmetric spiral dynamo forcing to the idea of dynamo quenching within the spiral arms by postulating outflows from the WIM aligned with the arms, perhaps driven by star-formation activity. In addition both \cite{Chamandy:2013aa,Chamandy:2015aa} and \cite{Moss:2015aa} consider the result of making the spiral forcing time-dependent with the idea of simulating the transient spiral structures seen in flocculent galaxies. In particular, \citet[][fig.~9]{Moss:2015aa} demonstrate that taking an axisymmetric dynamo-generated field (their fig.~9b) and suddenly introducing a co-rotating spiral pattern of high magnetic diffusivity (perhaps due to enhanced turbulence caused by star formation in the arms) can quickly (within 100 Myr) destroy the field in the arms (their fig.~9c), leaving only an inter-arm spiral magnetic pattern. This could have some relevance to what is seen in flocculent galaxies.}

\subsection{A simple model for spiral patterns}
We have argued that the magnetic structures we observe through polarization structures at $\lambda \approx 6$\,cm may come not from the centre of the disc plane, but from the higher parts of the WIM layer ($z \sim 300 - 1,000$ pc), well--separated from the disc regions ($ -100$ pc $ \lesssim z \lesssim 100$ pc) where the cool gas, molecular clouds and star formation reside. Given also the arguments of the previous subsection this suggests that we should now look at other possible interpretations. We note in particular that in most observed galaxies the ordered fields are strongest in the regions {\em between} the spiral arms. In the barred spiral M83, \cite{Frick:2016aa} also note that the polarized arms seem to be more or less independent of the spiral patterns seen in other tracers. They find that one of the main magnetic arms is displaced from the gaseous arm (as found in other galaxies) while the other main arm overlaps a gaseous arm. (We consider the exception M51 below.) It is worth noting that in most of these galaxies the spiral structure seems to be flocculent, so that it is most probably generated by internal self--gravitational instabilities within the stellar disc rather than by strong external tidal actions. This implies that the the spiral structures are short-lived ($\sim 100$Myr) and that the WIM gas flow relative to them must be very subsonic, i.e. essentially zero. This suggests a simple model (in line with the ideas proposed by \citealt{Moss:2015aa} outlined above). 

Suppose that the ordered galactic field is approximately an axisymmetric spiral, generated somewhere near the interaction zone or shear layer between the WIM and the halo, as discussed above. Then the pitch angles of the magnetic field and of the optical arms are not strongly related (as observed). Suppose that the main effect of the optical arms  (in the region $|z| < 100$ pc) on the region where the field resides ($|z| \sim 300$ pc) is to disturb it via outflows and feedback from the young stars (winds, supernovae) which form at low $z$ within and around the arm. Such outflows are likely to produce strong random motions at higher latitudes in the WIM. If so, then in the regions of the WIM  above the optical spiral arms the magnetic field structure would become disordered, and the polarized emission would be diminished. This simple idea seems to satisfy most of the observational constraints. 

For M51 the strength of the tidal interaction is sufficient even to shock the WIM ($10^4$ K gas -- \citealt{Dobbs:2010ab}). Thus here we might expect the strong shear present in the oblique shocks in the WIM to give rise to strongly ordered fields at the inner edges of the spiral arms \citep{Beck:2016aa}. It is sometimes argued instead \citep[e.g.][]{Mulcahy:2017aa} that the smoothness of the magnetic pitch angle is evidence for dynamo action because the large-scale field is built up over several galactic rotations and is not strongly affected by local features in the arms. However this argument would imply that dynamo action is too slow to reinstate ordered fields downstream of the arms. \cite{Patrikeev:2006aa} show that the magnetic-field lines up with the pitch angle of the CO spiral arm at the location of the arm, but changes inbetween the arms, consistent with our discussion.

\section{Discussion \& Conclusions}
\label{sec:discussion}
In this paper we suggest an alternative hypothesis to the standard view that the large--scale magnetic fields seen in spiral disc galaxies through observations of the polarization of radio synchrotron emission are generated by dynamo action within the galactic disc.

We argue instead that because of Faraday depolarization from the high densities close to the disc plane  ($|z| < 100$\,pc) the bulk of the polarized emission that we see comes not from within this region, but, rather, from closer to the interface between the warm interstellar medium and the halo gas at heights above the disc plane $|z| \sim 300-500$\,pc. We consider a simplified model in which the galaxy is initially threaded by a mainly poloidal field \citep[see also][]{Henriksen:2016aa}. \cite{Ideguchi:2017aa} consider a similar setup, and include in their analysis the effects of a turbulent field. We have \note{proposed \citep[see also][]{Henriksen:2016aa}} that the expected (and observed) vertical shear between the WIM and the halo can then produce a spiral magnetic field structure, similar to that observed. The shear between the WIM and the halo causes angular momentum transport, which winds the halo outwards and gives the X--shaped polarization structures observed in edge--on galaxies. These motions also induce a helical structure in the poloidal field which can explain the tendency for stronger Faraday depolarization seen at the receding end of the galaxy major axis, when observed at sufficient inclination angle.

It is found \citep{Beck:2016aa} that in galaxies with optical spiral arms, the mostly strongly ordered fields are generally between the optical arms, but that nevertheless, the pitch angles of the interarm magnetic field and of the optical spiral arms do not coincide. We suggest that this comes about because the pitch angles are generated by distinct physical processes. The magnetic field angle is generated primarily by the shear between the WIM and the halo, whereas the optical arms are generated by gravitational processes in the plane of the disc.  In addition we speculate that turbulent motions propagated into the WIM by stellar feedback (star formation, Type II supernovae) within the optical arms lead to less organised field, and so lower observed polarization, in those regions of the WIM above the optical arms. We have noted that in general the WIM does not feel strongly (if at all) the gaseous shock associated with passage through the optical arms, and that in flocculent galaxies there  is in any case essentially no velocity difference between the gas and the arms. However, in some galaxies (e.g. M51) where the arms are generated by a strong tidal interaction with a galactic neighbour (or neighbours) the WIM can also experience a shock as the arms pass by, and that in these galaxies there may be order in the magnetic field generated by such a passage.

We note that these ideas are mainly speculative at this point. More detailed cosmological simulations \citep[e.g.][]{Pakmor:2017aa}, along with higher resolution simulations of the galaxy itself, will be required to establish their degree of credibility.

\section*{Acknowledgments}
We thank the referee for a detailed and helpful report, and Clare Dobbs, Andrew Fletcher, and Dick Henriksen for useful comments on the manuscript. CJN is supported by the Science and Technology Facilities Council (STFC) (grant number ST/M005917/1). CJN was supported in part by the National Science Foundation under Grant No. NSF PHY-1125915. The Theoretical Astrophysics Group at the University of Leicester is supported by an STFC Consolidated Grant. TOH acknowledges support from the Swiss National Science Foundation grant number 200020\_162930. This work used the DiRAC {\it Complexity} system, operated by the University of Leicester IT Services, which forms part of the STFC DiRAC HPC Facility (\url{http://www.dirac.ac.uk}). This equipment is funded by BIS National E-Infrastructure capital grant ST/K000373/1 and STFC DiRAC Operations grant ST/K0003259/1. DiRAC is part of the UK National E-Infrastructure. We used {\sc splash} \citep{Price:2007aa} for the visualization.

\bibliographystyle{mnras}
\bibliography{nixon}

%%%%%%%%%%%%%%%%% APPENDICES %%%%%%%%%%%%%%%%%%%%%

%\appendix

\bsp
\label{lastpage}
\end{document}